# Beyond the Physics of Logic:
## Aspects of Transcendental Materialism or URAM in a Modern View

Rainer E. Zimmermann

IAG Philosophische Grundlagenprobleme,
FB 1, UGH, Nora-Platiel-Str. 1, D – 34127 Kassel /
Clare Hall, UK – Cambridge CB3 9AL[1] /
Lehrgebiet Philosophie, FB 13 AW, FH,
Lothstr. 34, D – 80335 Muenchen[2]
e-mail: pd00108@mail.lrz-muenchen.de

## Abstract

It has been shown at other occasions [1] that recent results of modern physics can be used to shed some more light onto the foundations of the world, provided the actual task of philosophy is being re-interpreted in terms of a theory which is following up the results of science rather than laying the grounds for the latter, contrary to what the original intention of Aristotelian *prima philosophia* would imply. As it turns out, the interpretation of the main results of present research dealing with aspects of quantum information theory and quantum gravity, respectively, as well as with self-organized criticality, suffices to re-construct a large class of phenomena not only within the field of physics proper, but also within chemistry, biology, and even the social sciences. As seen under a *philosophical* perspective, it can be shown that the general conceptualization of such a unified view of the world has been prepared on a long line of thought which begins with the Greek Stoá and leads up to the theories of Spinoza, Schelling and Bloch as some of its representatives, eventually showing up in a somewhat modified form in what can be called *transcendental materialism* today. [2] As seen under the *physical* perspective however, it can be demonstrated how a philosophy re-interpreted in the above-mentioned sense can unfold a heuristic potential which is able to produce guidelines of orientation as to deciding about competing concepts in physics. [3] On the other hand, such a philosophy can also hint towards what science cannot deal with at all, pointing to what is beyond the scientifically treatable. We show that questions of ultimate reality and meaning can only be answered within a framework of speculative philosophy which is rooted though in what sceptic philosophy is able to derive from scientific results. Hence, *cataleptic phantasy* is what is asked for, but this is not an arbitrary technique of imagination, but rather something whose nucleus is that sort of logic which itself is being produced by the underlying physics.

---

[1] Permanent Addresses.
[2] Present Address.

# 1   Philosophy & Science

Nothing is more difficult to mediate than the totality of mediation itself. Namely that which is encompassing and expressing the worldly (welthaft) existing as a unity which in turn implies that there is only one world for us in one piece (a holistic world that is) of which humans and all what they perceive, is a part. To be more precise: It is a produced fragment which is producing and re-producing itself all the time. All this despite the phenomenological variety of what can be concretely found within worldly (welthafter) *praxis* which appears to us very often as an irritating, disparate manifold of an amazing richness of form. In this sense, everything is indeed related to everything else: the physical foundations of the cosmology of our Universe to biological evolution, the mathematical modelling of the world to the aesthetical forms of expression in the arts, as well as the everyday life within the social process to the formation of structure within non-living nature. However, to merely accept this fact (which is difficult enough for its own part) means nothing but to support a somewhat tautological, if not even trivial, insight. What we need instead is the explicit derivation of the one from the other, i.e., a methodologically consistent, systematic illumination of the totality which establishes the interrelationships among all these different perspectives of the world which are themselves produced by the manifold of research fields – in turn showing up itself as a relationship between derivation and mediation at the same time.

Basically, it is only philosophy which uses a suitable language adequate to thematize this complex fact. This is so because the sciences (being primarily *single* sciences aiming at a unique field of research) are restricted *qua definitione* to a specific section of this totality. This is also true for aesthetics as the theory of the arts. On the other hand, the common language of everyday life (which indeed aims towards that totality) is by far too unclear and unprecise so that it is not able to open relevant insight into the worldly (welthafte) totality. What we need therefore, is a language which is situated somehow „within the middle" of all of these, and it is philosophy only which can offer such a language. This is mainly so because philosophy can be visualized as a „science of totality" from the outset (as Hans Heinz Holz has formulated a long while ago).

Unfortunately, this fact has been forgotten for long as far as main line philosophy is being concerned. In particular, the more recent philosophy of the 20[th] century, after world war II, has more or less completely suppressed the line of thinking which is mainly based on this fact, essentially in favour of a self-centring discussion of philosophy itself, taking its label for its contents. This is due to the neo-Kantian effort dating back to the 19[th] century, to eventually esta-blish a new, formal unity of the sciences, philosophy, the arts, and religion. Although this mainly apologetic enterprise has been subject to failure from the beginning on, it is nevertheless determining European (and in particular German) philosophy still today by its two basic components of activities: the history of philosophy on the one hand and the foundations of concepts on the other. Concrete *praxis* then, shows up only in terms of a developmental history of thinking, divorced from any practical relationship to worldly (welthafte) and also empirical existence – being enshrined in a kind of „pure motion of thoughts" which has lost already sight of the world of everyday life, of the socially mediated and politically active human being. Hence, the history of philosophy is very often mixed up with philosophy proper, the practise of *doing* history with actual philosophizing. And concentrating on the motion of thoughts means to actually forget about what these thoughts actually do reflect at all – thinking all the time, uncorrectly though, that the latter would not be important, once the mechanism of relexion itself would have been uncovered. This view is also supported, though from another perspective, by analytic philosophy which dominates the Anglo-Saxon region of languages. Here also, the systematization of thinking is more important than that about which thinking actually thinks. Moreover, it is assumed, also uncorrectly, that what is being thought can always be said, thus denying the unconscious and the necessity for any hermeneutic which is surpassing formal logic.

This is how philosophy leaves not only the conception of critically reflecting *praxis*, but is also leaving *itself*, because the former basically expresses the latter's nucleus of activity. It is not very astonishing therefore, that more and more often, philosophers are not able to contribute anything new to ongoing discussions of public life, not because they would not have available the appropriate inventory of concepts, but rather because they would not even recognize the problematic relevance of all these aspects of practical life. At the same time, this ideological tendency serves also a much more personal interest, namely one which aims at the securing of an isolated field of discourse which is exclusively reserved for the philosophical discourse alone. This has become very obvious in the recently emerged field of „cognitive science", where each explicitly materialistic model is being rejected from the outset by this „philosophical lobby", based on the false premise of indicating that this would be nothing than late relics of the mechanical materialism of the 19[th] century, and a primarily reductionistic effort with the objective of eventually taking over philosophy in merely scientific terms. However, keeping to this false premise means to forget about the fact that a philosophy which can be understood in terms of a dialectical and transcendental materialism visualized within a modern perspective is nothing but the result of critically reflecting recently gained insight in the various fields of science. In other words: only such an approach can be called *practical* in the first place. However, in order to really understand this idea, it is necessary to agree to two premises which so far have *not* been part of philosophical reasoning very frequently: 1) philosophy has to be understood as one which is dealing with the critical reflexion of what can be found within the world, and it is following up therefore the results of science rather than laying the grounds for them, 2) philosophy has to be understood as one which is orienting itself according to what science has found out about the world, in the first place, due to the actual horizon of knowledge of a given epoch. This is so because the sciences are „in charge" of the single sectors of worldly (welthaften) actuality, philosophy however, is „in charge" of the *totality* of all these sectors. Hence, in being oriented with a view to the sciences, philosophy is not at all duplicating the latter's work. Contrary to that, philosophizing means primarily to critically reflect each single sector of the uncovered worldly (Welthaften) with a view to the global interrelationships among all possible sectors, to ask for necessary and sufficient conditions for the main structural aspects of single sectors, and for possible alternatives. In this sense, philosophical reflexion gains more and more a *heuristic* connotation: The unification of the scientifically (necessary) segmented actuality is no self-purpose therefore (not a *l'Art pour l'Art*), in very much the same way as the arts are never self-purpose: On the contrary, the important point is to show, within all what is empirically observable, what *is not more* subject of science, as Theunissen has argued.

Only if philosophy is actually being visualized under this new perspective, will it be able to offer relevant insight and guidelines for an ethically adequate behaviour within the worldly (welthaften) *praxis* we do observe. Obviously, this viewpoint secures the relevance of what is going on in the sciences at the present time. There is no question of eventually falling back to $19^{th}$ century knowledge. Nevertheless, it cannot be disputed that the line of thinking on which we have based here this new viewpoint as one of its results, originates in the thinking of the German idealistic philosophers (especially the Tübingen group: Hölderlin, Schelling, Hegel), but also in the thinking of Feuerbach and Marx. The reconstruction of this line which in turn is based itself on earlier roots at least dating back as far as to Spinoza, has been topic then, in its modern version within $20^{th}$ century philosophy, in the conceptions of Bloch and Sartre, both different, but nevertheless also very similar in their detailed structure, being able in turn to give for us a relevant frame of orientation for a future philosophy still to come. It is the objective of this present paper to stress this point.

Hence, the important aspect is not to get rid of the historic genesis of thinking, and we can still learn from our predecessors. However, we cannot stop at this historical sight. We have to ask further: What shall we do with it? Namely given what is our present knowledge at the time. In this sense, *ethics is essentially knowledge* as I have formulated at an earlier occasion. [4] If this is the case, then we have to start with the knowledge. However, knowledge is always a mediated, communicated and communicable, thus linguistically expressable one. Hence, in order to arrive at the goal of a philosophy which can be visualized as one which is „charged" with heuristic potential, and which can give a feedback to the sciences again, after having critically reflected what is being known at a given epoch, we have to clarify three basic aspects: 1) the detailed nature of the afo-rementioned mediating totality, 2) the re-construction of a new concept of *praxis*, 3) the illumination of the relationship between this new concept of *praxis* and a similarly adequate concept of theory.

In the following we discuss these three aspects in terms of an example dealing with the fundamental categories which determine human thinking decisively from the beginning on: *space and time*. We will understand that it is physics itself which „pre-defines" these concepts which for us, are common everyday concepts at the same time, and which structures them in mathematical terms. However, in doing so, physics also demonstrates that these concepts turn out to be genetically produced concepts, and hence *they cannot be fundamental*. What we have to accept therefore, in order to dissolve this apparent contradiction, is that space and time are nothing but complex compositions of signs with which we characterize central aspects of the world we perceive, for giving a fixed basis to our own orientation. In this sense, these concepts underly a principle of selection which is very similar to the respective principle we use in biology. In other words: Humans show up as biologically produced entities whose mode of being is thinking, but who are also able to reproduce and produce aspects of their world (and themselves). Hence, we can think of humans as a *produced mode of production*. And it is this mode of production which is determining human experience decisively.

## 2   The Logic of Space and Time

There is hardly anything which would be more familiar to human beings than the characterization of the world in terms of the fundamental categories of space and time. It is time which fills the space of everyday life with meaning (to quote Bachtin) and nothing determines our everyday life more completely than (artificially defined) clock time. Concepts which are derived from this conceptualization of the world in terms of *space-time* determine themselves the structure of human language. Hence, communication itself is conceptually constituted in terms of space-time structure. A large class of languages incorporates a characteristic dichotomy of pairwise ordered concepts which model motion (including evolution) as a composition of changes per time distance. This dichotomy is also mapped onto the polar relationship of noun and verb which symbolically represents the contradiction between rest and motion. In fact, it is the grammar of the stoician Zeno dominating the whole European space of languages until today that has originally emerged from a detailed philosophizing about the relationship between space and time. Usually, the categories of space and time are being „corporalized" as a kind of container *in which* worldly (welthafte) processuality *is visualized of being happening*. Space and time provide a kind of „stage" on which the world *takes place*. World as theatre, as stage of life, as comedy is a collection of established literary *topoi*. In Kantian terms, space and time are the forms of conception (Anschauungsformen) under which the world is being taken into sight.

What is it then that we call *world*, and how can we apply these categories of space and time to this world? Basically, the world is visualized as the set of all that has been perceived – despite the fact that soon somewhat interpolated and extrapolated extensions creep into this conception, rarely noticed in everyday context, mainly due to the fact that we do not only cognitively process what we have perceived, but that we also *communicate* it all the time. Indeed: We can process the perceived at all, just *because* we have acquired a certain expectational intention (or anticipation) produced by our communicating in the first place. From our first days on we acquire techniques of an inventory which is there at hand for us. Socialization is mainly the learning of utilizing these techniques. However, once we would like to find a foundation for them, we have to draw from a lore which is present in our experience, an experience which is subject to the same subjective socialization so that our actions within daily life are based on a reflexion which can at most offer preliminary results. Actions (which always display behaviour towards Others and the environment) are limited therefore in their practical scope, effectivity and adequacy (in short: in their competence). Hence, a picture of the world which is primarily constituted from this somewhat incomplete information about the world is hardly able to provide us with guidelines for a really adequate behaviour (what ethics should actually do). What we can do at most is to find a moral consensus which refers us to a frame of reference as to what adequate behaviour could be. This frame of reference however cannot be otherwise than incomplete and therefore false. From there it is a long way to a reasonable ethics.

The basic reason for this divergence from what is aimed at is perception itself: This is obviously restricted from the beginning on such that *primary* perception (produced by the biological data of the sensoric input) is always fragmentary. So in *optical* terms we can only perceive a tiny section of the electromagnetic spectrum which we call light. Different sections (ultra-violet, ultra-red and gamma radiation) we can perceive at most as damage or in terms of organic defects, and they do not serve as controllable means of orientation. Similarly tiny is the *acoustic* section we perceive with our ears. And if we compare the optical and acoustic perceptions with the others (smelling, tasting, touching), they appear as *background* sensory perceptions rather than equal means of grasping the world. Hence, if we perceive only a tiny fraction of what there actually is, why should it be surprising at all that we have hardly access to our world *as it is really there*? Of course, measuring devices can help us to extend our means of perception. But this perception is not immediate, but mediated at best; it is *secondary* perception.

What we recognize is that humans gain their access to the world only in terms of mapping sensoric input to their brain. Instead of dealing with the „real thing" directly, humans achieve their access only by a two-step transformation of what there actually is: by a first step of transforming what there is into sensoric input data, by a second step of translating the incoming information to data which can be interpreted (determined by what is saved in the brain and is usually called experience). But even worse: The successive registration of input data is not even synchronous with the „picture" which is being composed then as interpretation of these data. This is because the brain releases a given information only when all data have arrived which correspond to that information. As we can clearly understand, the time data need to arrive at the appropriate brain section is different according to their nature: Obviously, optical data have to choose another lane for being transported to the brain than acoustic data have. The brain collects them into a single time window (of milliseconds). If this window is closed in the end, the brain presents us the corresponding information as a picture or a series of pictures similar to a movie. Strictly seen, we can say that the brain „composes" a movie representing the incoming information and „plays" it to us. The movie director is partially the biological „hardware" which fixes the causal frame of perception, partially the already saved „software" (which in turn consists of the internal „programmes" steering the brain itself, and of information acquired and composed in the past called experience).

Do not worry about our describing these aspects in terms of computer language. In fact, what we have done so far is not only to give a description of what is known about perception today, but we have also demonstrated that descriptions themselves, that is namings and connections between namings, can only be thought in terms of *metaphorizations*. Only a consensually accepted metaphor is able to express something which generates maximal understanding. And because many people deal by now with both movies and computers, it is very likely that metaphors from that domain, though obviously time-dependent as they are, can improve the understanding of the concepts underlying the ongoing discussion. In other words: Nobody would have been able to deal with the metaphors chosen here some hundred-fifty years ago. But because *today*, we visualize consciousness in terms of an information processing system, we can use a computer metaphor, because *for us today*, the computer is the leading paradigm. But this does not tell us how things really *are*, what we simply do is *to model* them according to our present leading paradigms which serve as foundations of communication. Hence, never is what we call *world* directly accessible to us. Everything is always mediated by a mapping procedure. In other words: According to what we perceive do we model the world and produce a *picture of the world*. In fact, we *produce* our world by means of this mapping procedure. (But note that the procedure we use is itself an outcome of the process we try to model!)

In philosophical terms, nevertheless, we assume the *realistic* viewpoint: We assume that exterior to our thinking and independent of ourselves there *actually is something*. But *how* what there is *actually is in truth*, we cannot find out at all. Because only a tiny fraction of what there is can be perceived. Hence, we can decide: either we call what we can model according to our cognitively processed perception *our world*. Then what exists *really* is exterior to our world, in particular beyond space, time and matter. Then we have a multitude of worlds, to assume anything else would be unjustifiable anthropomorphism. – Or, we call what there is outside, *the world*. Then our world is nothing but a mediated *partial aspect* of the totality which is this world. Hence, mediation is due then to the biological characteristica of humans. Again, there is a multitude of *perspectives* then under which the world can be taken into sight. Both cases have in common that there is one region about which we can know something, and another region about which we cannot know anything. And in both cases what we call „fundamental categories" is not really fundamental.

To philosophize about the one region is subject to *sceptic* philosophy, about the other is subject to *speculative* philosophy. But speculation does *not mean* arbitrary imagination: On the contrary, *speculation* is restricted with a view to the framework which is prescribed in terms of the sceptic approach reflecting what we already know about the world. That is to say, whatever speculation we choose, we have to arrive at least at the results of knowledge obtained about our world as far as it is accessible in empirical terms. This is a minimal compatibility condition we have to satisfy in any case. Even more: Although what we call fundamental categories turns out to be more than questionable, whatever model we actually choose, we have to end up with the result that what we do perceive must be compatible with what we *cannot* perceive, because as *part or aspect* of totality we always remain part or aspect of *totality*.

And this confronts us with another problem: If we accept the aforementioned we have to differ in future between two perspectives or two levels of description. What we can know and what falls therefore into the domain of sceptical philosophy, can be expressed in terms of language, in particular in terms of propositions. As we have seen already, language itself is structured in spatial and temporal terms, and reflects therefore the fundamental categories which serve as framework for our thinking. This is different however in the case of the second (in principle more fundamental) perspective: In that case we have to deal with a region which is not itself constituted in spatial and temporal terms. Nevertheless, for its description we have to use the same language which is based on the fundamental categories which „have nothing to say" about this region. Obviously, this must be very difficult. But we recognize now the advantage of our language being constituted such as to act basically by means of metaphors: only then do we have a possibility at all to signify things which lie outside our inventory of signs. Despite of this we can at least approximate meanings which might eventually gain practical relevance.

The consequences are obvious: Scientific modelling relies always on the sceptical aspect. In terms of a spatio-temporal perspective it is not difficult therefore, to talk about „evolution". We can state the following: Humans are biological entities which have emerged from the evolutive differentiation of structures within the interplay of selection and mutation on this very planet. However, as can be clearly recognized, this is only true according to the sceptical perspective. Because, according to the speculative perspective, there is no evolution, because there is no spatio-temporal processuality. In other words: Human consciousness is structured in such a way that its modelling leads to reasonable results, provided we model them according to the fundamental categories of space and time. This enterprise is nothing but practical. In this sense, a speculative foundation of the same model is rather *unpractical*. But it is the only way to eventually achieve something like a „true" foundation. Because, as Schelling already knew, the foundation of something is always outside of what it is foundation to: „ ... foundation is against that to which it is foundation, non-being." [5] Hence, the first form of foundation is nothing but the regression onto the origin: Everything which actually is, can at best be *reduced* to its beginning. (Schelling called this „mechanism".) This foundation however is that of praxis. The second form makes possible to *derive* what is out of its foundation. The possible is being unfolded by going progressively forward to what there has become. (Schelling called this „organism".) Both these movements are not identical: The regressive component displays what there *actually has become*, the progressive component displays what there *could have become*. The philosophical insight into the foundation of totality is based exactly on that discrepancy between the two movements. In this sense, foundation means to uncover possibility and actuality, to construct the boundary between them, and to talk about their mutual relationship. We note two basic results: First of all, the foundation of being *is not nothing(ness)*, but it is *something*; it is what *is not*, but *can be*. It is possibility. Nothing(ness) however is what *is not*, but also *cannot be*. It is the impossible. If something has become actual, then it must have been possible in the first place. The inverse is not true. [6] (And hence, there cannot be a *creatio ex nihilo*.) On the other hand, this is rather a methodological insight, speculative philosophy goes onto progression. (Schelling called it „positive".) Sceptical philosophy goes onto regression. (Schelling called it „negative".) In manifold variations this contradiction is always intrinsic and shows up as contradiction between synthesis and analysis, and between deduction and induction. (At the time of Schelling, people referred to the contradiction between rationalism and empiricism. [7] )

But note: Philosophy cannot stay with this contradiction alone. Because every methodological dualism is itself perspectively determined. In reality, regression is always limited from the outset, because it is sceptical. And although the sceptical perspective refers back to the speculative one, as to its „natural" contradiction, this contradiction itself is somewhat artificial again, because it is actually being produced by the restriction in terms of perspectives in the first place. Strictly spoken, this contradiction is itself negative. Consequently, Schelling tried to derive a „true principle of being" which he called *potentia ultima*. This potential was thought as indicator for a „reality before all possibility". [8] He attempted this approach by means of a basically idealistic philosophy. Nevertheless, his merit is mainly to visualize the problem in systematic terms from the beginning on such that method and system fall into one. Although Schelling falls back behind Spinoza who aims at existence rather than at essence, and also introduces important *materialistic* connotations into his philosophy (because for the latter, the attribute of substance, is in the end space itself), he recognizes the *organicity* of systems: „The system must have a principle which is constituted in itself and by itself, which reproduces itself in every part of the whole; it must be organic: one must be determined by all, and all by one: it must not exclude anything, not unilaterally subordinate or even suppress anything." [9] In this formulation, Schelling expresses as early as 1810 what today turns out to be a central aspect within the context of modern research: *reproduction*. If the whole system reproduces itself, then its subsystems are being produced. However we recognize the critical point here: *Reproduction* can remain static and is not automatically referred to a concept of motion or of dynamical process. *Production*, on the contrary, points always to a dynamical activity, because something *acts* upon something else. Hence, self-reproduction can be visualized as permanent, instrinsically resting, processing of itself by itself. But production asks for history. It asks for spatio-temporal constitution.

Hence, we are back to perspectivity: The excellence of Spinoza's approach is in the constitution of the attribute (which is essentially one). The attribute of extension (res extensa) has to be visualized as a whole which acts onto itself by self-diffentiating into parts. It is *not constituted* by parts. [10] Also, for Spinoza, it is only the imagination which goes onto parts. In reality, matter is only one. Its partition can only be observed *modaliter*, but not *realiter*. Space is the principle of matter as principle of order. But space *is not really there*: it permanently *processes* itself. In so far it satisfies Schelling's organicity condition. And what Spinoza calls *attribute*, Schelling would call *system*. The world as it is understood in terms of a system, this is nature as worldly (welthafte) totality. If we take both of them, Spinoza and Schelling, together, we recognize as the principle of the self-organizing worldly (welthaften) being *producing space* which reproduces itself in producing its parts.

## 3   The Physics of Logic

Now what about the viewpoint of physics which is the appropriate science „in charge" of the problems discussed here? It is surprising to note that modern scientific research is dealing with not less than five aspects of what we have discussed so far: 1) with the relationship between substance and attribute (if we stay with the Spinozist terminology for a while), 2) with the onto-epistemic mediation of what there is with what there is being known, 3) with a universal principle of selection governing the unfolding of structures in nature, 4) with the detailed ordering of a hierarchically structured derivation of the evolution proper, 5) with the methodological transition from logic to hermeneutic. I have discussed technical details of this at other places (see the above references). We simply concentrate here on the essential results and ask what they actually do mean with respect to the problems discussed here.

The main problem of physics today is the fact that apparently, physics takes place in two disjoint domains (or worlds) each of which claim to represent the one and only world: On the one hand, we have Einstein's theory of relativity, describing *macroscopically* a world which can be visualized as *space-time*, i.e., a four-dimensional space with a *metric* (an invariant measure of distance between points) and a specific *signature* of this metric (meaning that the sign of the temporal coordinate is different from the other signs). – On the other hand, we have quantum theory, describing *microscopically* a space which is not really a space of concrete observables, but an abstract state space (Hilbert space) which essentially describes the set of all possible states of a physical system. This space has a large number of dimensions (interesting cases are actually infinite-dimensional), and its signature does not indicate any sign differences. In other words: two completely different descriptions of one and the same world. John Baez has recently called this a „deeply schizophrenic world view of physics". [11] Hence, relatively early, the idea of „pre-geometry": If we are not able to unify both these disjoint theories *within* the world, it might be possible to do so *outside* the world. So the basic idea was to introduce an abstract mathematical structure from which it would be possible to actually *derive* space and time (and matter), the contradiction between relativity and quantum theory included. First approaches are connected with the names of John Wheeler, Roger Penrose, and Stephen Hawking. Today, this attempt is dominated by two competing theories: the *theory of superstrings* (or M theory recently) and the *theory of loops*. The former visualizes the elementary buildings blocs of the world on the fundamental level as string-like entities which vibrate in various oscillating modes (however in a space of up to eleven dimensions) whose oscillations produce „tones" (similar to a violine) representing elementary particles. The latter is not so different in starting from microscopic loops which essentially describe tiny inherent curvatures and can be visualized as knotted strings. There is convergence, but both of these theories have their disadvantages: The string people suffer, because they have to start from a background space which is very similar to that of special relativity theory – although their theory shall derive space and time in the first place! The loop people suffer, because the technical difficulties of their theory make it unlikely to achieve many empirical confirmations very quickly. They can correctly claim however, that they start from a purely combinatorial structure without any reference to space or time (which is very much in the sense of Einstein's basic idea).

At the end of the sixties (of the last century) Roger Penrose has proposed a sort of derivation which now, thirty years later, turns out to be of relevance for the loop theory. As „skeleton" of the world he chooses an abstract network whose „knots" have the property of *spin* (individual angular momentum) only. They permanently exchange spin according to the conservation of angular momentum. Hence, this *spin network* is a fluctuating web which expresses nothing but this permanent exchange of spin numbers among knots. It is purely combinatorial and thus underlies the observable space. In fact, physical space can be thought of as superposition or „condensation" of many „layers" of this network. There is, together with the spin network, a dual web consisting of a *triangulation*, a covering of the network by triangles such that their centre of mass is identical with one of the network's knots. This can be visualized as minimal quantization of space in the following sense: Because the covering de-composes „proto-space" into elementary portions described by the length of the triangle sides, one can define an intrinsic quantization of space volume and space surface which corresponds to the configuration of spin numbers. Hence, re-arrangements of spin numbers cause re-arrangements of quantizations; only the duality relation remains conserved. This describes the microscopic domain.

On the macroscopic domain which is the observable level of the (classical) physical world, we have the processes whose dynamics represents the formation of structure (i.e. the production of new structures). Under the perspective of a classical, macroscopic observer, formation of structure means basically change of space topology. The phenomenological space is three-dimensional. It can be understood as the boundary of four-dimensional space-time. Hence, a change in the components of this space implies a change of its boundary. In mathematical terms, the formation of structure in this sense, is a mapping from an old space configuration to a new one. Such mappings are called *cobordisms*, because they map spaces onto each other which are *cobordant*.[3] The important point is that cobordisms as describing a change of structure are not themselves functions, but spaces! Which has the following consequence: A cobordism describes the temporal change of a space topology which is expressed in terms of the complete space-time, not only by the time coordinate, as is the usual case for classical physics. Hence, time shows up here as a quotient of change which is being integrated into the form of space itself and which is not separable from it. The reason for humans nevertheless „observing" time as being separated from space is simply that we are obliged to perceive everything in a sequential manner (one after the other). We cannot perceive everything at the same time. Hence the old proverb: that time is what prevents everything from happening all at once. We can read this as a direct confirmation of the fact that time turns out to be a human property of perception, but nothing that would be „really" there independent of humans.

We note two important consequences: On the one hand, it can be shown that there is an explicit correspondence between each three-dimensional space component and a Hilbert space such that for each cobordism on the macroscopic level we have a microscopic change of state. On the other hand, cobordisms satisfy axioms of (mathematical) categories. They are essentially sets of objects and mappings among objects (morphisms) which obey two axioms (of the existence of identities and compositions). A fundamental category in physics is that of Hilbert spaces (Hilb) whose objects are the spaces themselves, whose morphisms are linear operators among them. In this sense, objects of the category of cobordisms (nCob) are (n-1)-dimensional spaces, the morphisms among them are n-dimensional cobordisms. Mappings among categories are called „functors". They map objects to objects and morphisms to morphisms.

---

[3] As the name implies, two spaces are said to be cobordant, if there is another space of one dimension more such that its boundary is the disjoint union of the other two spaces.

Because it is primarily the morphisms that determine the characteristics of a category, the theory of categories is mainly one which deals with relationships and interactions among objects and morphisms rather than with constitutive parts or elements in the fashion of set theory. Hence, as higher-dimensional analoga to sets, categories have their specific meaning for the evolutionary aspects of physics. The central point here is that a theory of everything (TOE) like quantum gravity shows up as a functor itself. This functor is the aforementioned mapping which represents the correspondence between the categories nCob and Hilb. It completely expresses the form of the theory and its contents.

It is comparatively easy now to recognize the basic insight of this referring us back to the original five points under discussion: The actual relationship between the fundamental level of physics and the classical level of physical phenomenology can be visualized as a formal analogue of the relationship between substance and attribute. The concrete world of daily perceptions shows up as a perspective restriction of the real world on the fundamental level. In particular, the formerly „fundamental" categories of space and time are uncovered as being derived categories instead. On the other hand, the representation of the theory and of what the theory itself represents turn out to be one and the same. In this sense, we can speak of „onto-epistemic" mediation (according to a concept introduced by Sandkühler). If the theory *is* the functor by which it is represented symbolically, then there is a systematic parallelism between thinking and what thinking is thinking about. This can be easily compared with the conception of Spinoza.[4] Even more: The mediation of the fundamental and classical levels of physics underlies itself the correspondence-producing action of the theory-functor. Hence, we can assume that there is a unique principle which universally steers the production of correspondences. Recently, several ideas have been presented in order to locate such a universal principle. [13]

Obviously, the problem of finding an adequate translation of abstract fluctuations in a spin network on the fundamental level into phenomenological processes which can be observed in the world is nothing else than the problem of finding a translation of the level of substance into the level of attribute. The actual foundation of our world is at stake. But there is one more point to note: Basically, logic itself can be represented by algebraic structures such that the explicit semantics implied by this logic is expressible in algebraic terms. The axiomatic structure of a category is nothing but such an algebraic structure. Hence, each category represents a concrete process, represents at the same time the mapping (representation) of this process and uncovers the logical structure of this representation. This threefold function of a (mathematical) category secures the consistency of the aforementioned fact that *the theory is the functor*. Part of the axiomatic underlying the category is the composition rule: If A, B and C are objects of a category, and f, g, h are morphisms of the form

$$f: A \to B, \; g: B \to C, \; h: A \to C,$$

then we shall have for the composition of the first two: $g \circ f = h$. Now note the following: The objects can actually mean complete *languages* (sets of objects = words and morphisms = correspondences which follow a given axiomatic structure = their logic = grammar). Assume that A stands for German, and B for English, and C for French (so that the chosen category is that of all possible languages). Then we clearly recognize here a crucial difference as compared to formal languages (such as mathematics): While in the case of the latter, translations must be always complete (so that the composition rule is strictly satisfied), in the case of the former, this is not possible. A German text which is translated into English first, and then into French will usually be different from a German text which is translated directly into French. In other words: $g \circ f \neq h$. Why is that? Because, in general, a translation between logics is always „exhausting" or not possible at all (null). The underlying axiomatic structure guarantees consistency. (Extraterrestric humans, if we meet them one day, would always be able to translate our mathematical representations into their (mathematical) language. Conversely, we would always be able to comprehend what they mean when talking about mathematics. Otherwise we would not have met them in the first place.) Not so in the case of everyday language: Contrary to formal languages, it has the task to map everything what there is, independent of strict, logical requirements. But why is it that we cannot express what is beyond logic in adequate terms? Because we lack the necessary information: There are many things which cannot be forged into any closed, consistent, propositional form. Hence, we can never completely know what is meant with what is said. What we actually mean depends on the individual context, and this is mainly a result of a singular socialization. (Hence, the real problem as to these extraterrestric people would be to find out something about their emotions which determines their mentality decisively – and our political behaviour.) In the case of everyday language not logic is at stake, but hermeneutic. This is not an art of knowing, but rather an art of guessing. (This does not mean that the latter would not have to rely on the former.) So logic is for hermeneutic the necessary, but not sufficient condition of understanding. Hence, hermeneutic can be visualized as *generalization of logic by incomplete information*. Logic turns out to be a special case of hermeneutic: It is valid, if translations are *path-independent* (in the sense of category theory). In general and complex situations however translations are *path-dependent*.

---

[4] Ethics 2p7: „Ordo et connexio idearum idem est ac ordo et connexio rerum." [12]

So we notice a surprising fact: that crucial aspects of the problems intuitively discussed before with a view to philosophical argumentation are already for a long time subject of scientific research. Only the language utilized and the objectives defined by the respective research are different. Looking more closely reveals however that this is not very surprising indeed, given the assumption we mentioned before: that humans are part of nature, a producing product of nature, to be more presise.

## Preliminary Conclusions

Coming back to our computer metaphor: The recent (and very successful) movie *Matrix* [14] has illustrated another aspect of what we have discussed so far. The movie describes a future in which humans are being exploited by intelligent machines which deceive them by an artificial world (whose software is called *matrix*) in order to control them. The problem of the protagonists (a small group of humans who was able to „de-couple" from the matrix) is to find a way of influencing (that is: re-programming) the matrix software. We will not discuss the various inconsistencies of the movie's plot here or the celebrated action scenes. (In fact, the movie is nothing but a recent version of an old idea.[5]) What we will do is to compare the basic idea with our topic here. Obviously, there is an „onto-epistemic" level of interpretation in that movie which implies the simple statement that nobody is able to determine his/her's ontological state. Because (given a „perfect simulation") there is no way to give a criterion for differing between simulation and real world (or concrete world *modaliter* and real world *realiter*). This is nothing but the problem of substance and attribute. But using here the computer metaphor (because we talk about actual simulation software), displays humans as subroutines of a higher-order matrix programme to which we have no perceptive access. As a part of a larger programme, we cannot have knowledge of detached parts of the same programme, unless by indirect speculation. But the latter would not eventually enable humans to gain actual insight into the matrix programme, but would rather serve the purpose of somehow improving the insight into our own programme. Which is far more modest. And it is perhaps this sort of modesty which would serve as a reasonable basis for any future ethics which is to be derived from the resulting relationship between sceptic and speculative aspects of our present model of the world.

In fact, this perspective, to visualize the world as a permanent computational process which is part of a larger whole, is very much compatible with what is going on in present research: So far, there are only few publications dealing explicitly with a combination of quantum gravity and quantum computation, but a number of important indications are already at hand. Indeed, classical physics could be visualized as one which emerges from a choice of transport channels processed quantum information on the fundamental level has to make. These channels would be the vertex lines of the spin network through which information is able to percolate (like coffee in a filter).

---

[5] This is going back to a novel by Daniel F. Galouye's of 1964 which has been adapted to a movie for German television as early as 1973 by Rainer Werner Fassbinder. There is another movie by Josef Rusnak with the same topic which is much nearer to the novel: The 13[th] Floor, 20[th] Century Fox, 1999.

But we realize that in order to answer the open questions, only an interdsiciplinary and mutually heuristic cooperation of both science and philosophy can produce deeper insight into the underlying fundamental structures. So far, such a cooperation has been publicly demanded but more often feared in practise. It is time now to take the next step.

## Acknowledgements

I thank for stimulating and illuminating discussions on the topic Mary Hesse (Cambridge, UK) and Richard Bell (Wooster/Cambridge, UK).